\documentclass[%
 reprint,
 amsmath,amssymb,
 aps,
 prd
]{revtex4-1}

\usepackage{graphicx}
\usepackage{dcolumn}
\usepackage{bm}
\usepackage{siunitx}

\DeclareSIUnit \h {\ensuremath{\mathit{h}}}
\DeclareSIUnit \parsec {pc}
\DeclareSIUnit \jy {Jy}
\DeclareSIUnit \arcmin {arcmin}
\DeclareSIUnit \year {yr}
\newcommand\mnras{MNRAS}
\newcommand\jcap{J.~Cosmology Astropart. Phys.}
\newcommand\procspie{Proc.~SPIE}
\newcommand\pasp{PASP}
\newcommand\aap{A\&A}

\begin{document}

\preprint{APS/123-QED}

\title{Constraining the Expansion History and Early Dark Energy with Line Intensity Mapping}

\author{Kirit~S.~Karkare}
\affiliation{%
Kavli Institute for Cosmological Physics, University of Chicago, Chicago, IL 60637, USA\\
Department of Physics, University of Chicago, Chicago, IL 60637, USA
}%
\email{kkarkare@kicp.uchicago.edu}

\author{Simeon Bird}
\affiliation{University of California, Riverside, Riverside, CA 92521, USA
}%

\date{\today}

\begin{abstract}
We consider the potential for line intensity mapping (IM) experiments to measure the baryon acoustic oscillations (BAO) from $3 < z < 6$.  This would constrain the expansion history in a redshift range that is currently unexplored.  We calculate the map depths that future IM experiments
targeting the CO(1-0) rotational transition line and [CII] ionized carbon
fine-structure line would need to achieve in order to measure the BAO. We find that near-future IM experiments could constrain the BAO scale to 5\% or better depending on CO/[CII] model amplitude. This measurement is at a precision that could make competitive constraints on models of early dark energy.
\end{abstract}

\pacs{Valid PACS appear here}
\maketitle


\section{\label{sec:intro}Introduction}

Measurements of the Hubble constant \cite{Riess16}, the Cosmic Microwave Background (CMB) \cite{planck16}, and the Baryon Acoustic Oscillations (BAO) \cite{Eisenstein05} have firmly established that the energy density of the Universe is today dominated not by matter but by dark energy, a component with an equation of state $w \approx -1$, leading to  accelerated expansion. However, the nature of this dark energy remains a mystery.

In order to elucidate this mystery, cosmologists make precise measurements of the expansion history through standard candles, such as supernovae~\cite{Betoule:2014frx}, and standard rulers, such as the BAO scale \cite{Eisenstein:2005}. The expansion history in turn constrains the properties of the dark energy equation of state, commonly parametrized as $w = w_0 + (1 - a) w_a$~\cite{Chevallier:2000qy}. BAO measurements have been made in several large samples of galaxies at different redshifts, including 6dFGS at $z=0.106$~\cite{Beutler:2011hx}, SDSS at $z=0.15 - 0.6$~\cite{Ross:2014qpa}, and DES yr 1 at $z=0.81$~\cite{desyr1}. At higher redshifts, the BAO are measured using the correlation of Lyman-$\alpha$ forest flux at $z = 1.5$ \cite{Ata2017}, $z=2.33$~\cite{Bautista:2017zgn} and $z=2.4$~\cite{Bourboux:2017cbm}. This range will also be probed by upcoming \SI{21}{\centi\meter} neutral hydrogen (HI) experiments such as CHIME \cite{bandura14} and HIRAX \cite{newburgh16}. Finally, at $z \approx 1100$ the BAO distance scale is anchored by the position of the first CMB peak \cite{planck16}.

At present, there is a redshift gap from $z=2.4$ to $z \approx 1100$ during which we have no firm measurement of the expansion history. Traditional galaxy surveys become sparse because few high-$z$ sources are bright enough to detect in sufficient numbers. Several neutral hydrogen experiments are targeting the Epoch of Reionization (EoR) at $z > 6$ (e.g. PAPER \cite{ali15}, MWA \cite{beardsley16}, HERA \cite{deboer17}), but none currently plan to measure intermediate redshifts. Consequently, our knowledge of the expansion history and thus the behavior of dark energy at this time is incomplete. In the standard model, the Universe is highly matter dominated in this redshift range. However, there are two important reasons to investigate filling this redshift gap.

First, a well-known minimal extension invokes a component of ``early dark energy'' at these redshifts \cite{Wetterich2004, Pettorino2013}. Early dark energy models posit that a sub-dominant (but non-negligible) fraction of the energy density of the Universe at $z \gg 1$ is made up of a scalar field with negative pressure. This early dark energy later decays---to matter or to a cosmological constant. A recent simple parametrization suggests that early dark energy has $w = -1$ until some critical redshift and thereafter has $w = 1$, the equation of state of a free scalar field \cite{Karwal2016}. Early dark energy models, which can be motivated from string axiverse models \cite{Kamionkowski2014}, provide a dynamical mechanism for dark energy and attempt to naturally explain the observed similarity between $\Omega_m^0$ and $\Omega_\Lambda^0$
\cite{Aubourg:2014, Linder:2015,  Shi:2016, Martins:2016, Caprini:2016}.

Second, the expansion rate measured from Lyman-$\alpha$ BAO is currently in moderate tension with the standard model \cite{Poulin2018}. If this tension is confirmed by other experiments, it will become imperative to measure the expansion rate at higher redshift and thus determine when it first appeared. In this paper, we will assess the ability of line intensity mapping measurements to fill this redshift gap by providing BAO measurements in the redshift range $z=3-6$.

A promising technique for probing large cosmological volumes is line intensity mapping (IM), which uses a relatively coarse beam to measure a spectral line integrated over many unresolved sources \cite{kovetz17}. IM is thus more analogous to CMB measurements than a traditional galaxy survey. Compared to galaxy surveys which require emission to be above a flux limit, IM measures all of the line-emitting sources. Since only enough resolution to resolve the fluctuations in large-scale structure is required, cosmological volumes can be surveyed much more quickly.  Redshifts are obtained through the frequency dependence, making the maps inherently three-dimensional.  Maps produced this way are capable of measuring large numbers of modes at various points in the history of the Universe, probing both cosmology and high-redshift astrophysics.

Several candidate lines for IM that emit in the early Universe are being explored. These include the \SI{21}{\centi\meter} spin-flip transition of HI \cite{madau97, pritchard12}, the ionized carbon [CII] fine structure transition \cite{gong12}, the rotational transitions of carbon monoxide (CO) \cite{righi08, lidz11}, and Lyman-$\alpha$ \cite{silva13}.

The CO $J = 1-0$ rotational transition line---hereafter CO(1-0)---traces dense molecular gas and star formation, both locally and in distant galaxies.  Emitting at a rest-frame \SI{115}{\giga\hertz}, by $z=0$ it is redshifted to the \si{\centi\meter}.  It is considered a prime target for IM \cite{gong11, carilli11, lidz11}, and indeed several experiments (e.g. COMAP \cite{li16}, Y.~T.~Lee Array \cite{ho09}) are currently targeting detections at $z \sim 2-3$.  There is also evidence for nonzero CO power at $2.3 < z < 3.3$ from the COPSS experiment \cite{keating16}.

Similarly, the \SI{158}{\micro\meter} fine-structure transition of ionized carbon---hereafter [CII]---is a promising IM candidate: it is generally the brightest emission line in star-forming galaxies, and has been observed in individual galaxy spectra out to $z > 5$ \cite{riechers14}.  Submillimeter spectrometers are now being built to measure [CII] fluctuations from $4 < z < 9$, including CONCERTO \cite{serra16} and TIME \cite{crites14}.  Ref.~\cite{pullen18} presented evidence for [CII] emission at $z \sim 2.6$ in \emph{Planck} cosmic infrared background maps cross-correlated with quasars and CMASS galaxies from SDSS.  

These tentative detections are encouraging for IM experiments.  Understanding the constraints that these measurements, or futuristic versions thereof, can place on the expansion history is therefore valuable.

In this paper, we determine the constraining power that IM experiments targeting CO(1-0) and [CII] could provide on the BAO scale at the high redshifts $3 < z < 6$.  Such measurements would constrain the expansion history at a previously-unexplored period in the history of the Universe and probe exotic models of dark energy.  While much previous work has explored the implications of these measurements for models of galaxy formation and high-redshift astrophysics \cite{li16, breysse16}, the potential for cosmology has received less attention \cite{fonseca17, dizgah18} (however, we note that Ref.~\cite{bull15, bull16} explored the constraints on $w$ achievable by \SI{21}{\centi\meter} IM with the SKA at $z < 3$). We calculate the noise performance that IM experiments would need to achieve to constrain BAO to percent-level precision. In the absence of complications such as continuum foregrounds and interloper emission lines, our results indicate that for many line emission models in the literature, realistic next-generation IM experiments could make \SI{5}{\percent} or better constraints on the acoustic scale at $z = 3-6$.  Measurements at this precision would provide competitive constraints on models of early dark energy.

\section{\label{sec:bao}Baryon Acoustic Oscillations with Line Intensity Mapping}

Emission lines targeted by IM experiments trace the underlying matter on large scales. Most of these IM tracers originate within galaxies, and thus are more clustered than the matter power spectrum, $P_{\delta \delta}(k,z)$. We parametrize this clustering change by a line-dependent bias $b(z) > 1$. The lines also have mean brightness temperatures $I(z)$ that
change with redshift---in the case of e.g. [CII] or CO, as more stars
form and metallicity increases, so does the temperature.  The clustering
power spectrum which allows us to constrain cosmology is then
\begin{equation}
P_{\text{clust}}(k,z) = b^2(z)I^2(z)P_{\delta \delta}(k,z).
\end{equation}

The total power spectrum measured by an IM survey also includes noise components:
\begin{equation}
P(k,z) = P_{\text{clust}}(k,z) + P_{\text{shot}}(z) + P_N(k,z),
\end{equation}
where $P_{\text{shot}}$ is a Poisson term
due to the discrete nature of the line-emitting galaxies and $P_N$ is
instrumental noise.  Note that $P_{\text{shot}}$ is constant in $k$.  For an
experiment with white noise, $P_N$ is also constant in $k$---however, due to the finite resolutions of the beam and spectrometer of any real instrument, the noise is inflated at high $k$ \footnote{The beam and spectral binning attenuate the measured power spectrum at high $k$.  This factor can be divided out at the expense of increasing the originally-flat noise.}.  An experiment with a beam FWHM $\theta_b$ and spectral resolution $\delta \nu$ has a transverse smoothing scale $\sigma_{\perp} = R(z)\theta_b / \sqrt{8\ln 2}$ and a parallel smoothing scale $\sigma_{\parallel} = c \delta_{\nu} (1+z)/ \left[ H(z)\nu_{\text{obs}} \right]$, where $R(z)$ is the comoving radial distance.  Letting $\mu$ be the cosine of the angle of the Fourier mode $k$ with respect to the line of sight, the noise in a spherically-averaged estimate of the power spectrum for a white noise level $P_{N0}$ becomes \cite{li16}
\begin{equation}
P_N(k) = P_{N0} e^{k^2 \sigma_{\perp}^2} \int_0^1 e^{\mu^2 k^2 \left(\sigma_{\parallel}^2-\sigma_{\perp}^2\right)} \ d\mu.
\end{equation}

Given a cosmological power spectrum, a noise spectrum, and a survey volume, the uncertainty on a measurement is given by
\begin{equation}
\label{eqn:sigma}
\sigma^2(k) = \frac{\left[ P_{\text{clust}}(k) + P_{\text{shot}} + P_N (k) \right]^2}{N_m(k)};
\end{equation}
$N_m$ is the number of modes in each survey bin. A bin centered at $k$ with width $\Delta k$ within a survey volume $V_s$ has $N_m = k^2 \Delta k V_s / 4\pi^2$. A survey with $j$ bins centered at $k_j$ in turn has a signal-to-noise ratio (SNR) for \emph{detection of the clustering power spectrum} of \cite{pullen13}
\begin{equation}
\label{eq:snr}
\text{SNR} = \sqrt{ \sum\limits_{j} \left( \frac{P_{\text{clust}}(k_j)}{\sigma(k_j)} \right)^2}.
\end{equation}

In this paper we are concerned not with detecting the power spectrum alone, but with constraining the expansion history using the BAO feature. The BAO, originating from sound waves in the primordial plasma, manifest as an enhancement in the real-space galaxy
correlation function at a characteristic distance of \SI{110}{\mega\parsec\per\h} or
equivalently a series of oscillations in the power spectrum around $k
\sim \SI{0.1}{\h\per\mega\parsec}$ with a characteristic scale of \SI{0.06}{\h\per\mega\parsec}.
We must thus compute the sensitivity not for the power spectrum, but for this enhancement. 
The amplitudes of the BAO peaks are small ($\sim 5\%$) compared to the
absolute power spectrum amplitude. However, it is the \emph{position} of
the BAO that contains cosmological information: since the
length scale in real space is known, the angular component of the feature
on the sky measures the angular diameter distance and the radial component measures $H(z)$ \cite{eisenstein98b}.  

For a given survey, noise level, and model for CO or [CII] emission, we ask what constraint on the BAO scale can be obtained from IM assuming a standard linear $\Lambda$CDM power spectrum.  
To answer this, we follow the approach outlined in Refs.~\cite{seo07, seo10}.  We allow the distance scale to change by parametrizing the power spectrum as
\begin{equation}
P_{\text{clust}}(k,z) = b^2(z)I^2(z)B(k)P_{\delta \delta}(k/\alpha) + A(k),
\end{equation}
where $\alpha$ dilates the scale of the BAO and $A(k)$ and $B(k)$ are smooth functions which equalize the broadband slopes and offsets between models with different $\alpha$.  For a survey with uncertainties given by Equation~\ref{eqn:sigma}, we assume a Gaussian likelihood and evaluate the posterior for $\alpha$ using a flat prior over $0.85 < \alpha < 1.15$ \footnote{The prior range was chosen to constrain the BAO to cosmologically-useful precision while allowing for the possibility of $\sim 5\%$ deviations as suggested by the Lyman-$\alpha$ BAO.  A flat prior makes our forecasts slightly more conservative than what we would obtain using a Gaussian prior centered at $\alpha = 1$.}.  The $1\sigma$ width of the posterior, $\sigma(\alpha)$, represents the survey's constraint on the acoustic scale.
Throughout this paper we report both $\sigma(\alpha)$ (in percent relative to a fiducial $\alpha=1$) and the SNR of the overall power spectrum detection to facilitate comparison to other work. 

\section{\label{sec:results}Results}

In this section we consider the survey volume and resolution element required by a hypothetical intensity mapping experiment in order to resolve the BAO in the sample variance dominated limit. We then calculate, for two surveys targeting the CO(1-0) and [CII] emission lines, the map noise required to reach various $\sigma(\alpha)$.  In Section~\ref{sec:future} we will use these noise levels to calculate the feasibility of detection using realistic near-future experiments.

For both CO and [CII] there is significant modeling uncertainty in the amplitude of the expected signal; for example, estimates for the amplitude of the CO clustering component vary by over two orders of magnitude (e.g. \cite{li16}).  So that our results are as general as possible, instead of choosing a fiducial model for each redshift, we evaluate detection significance for a wide range of model amplitudes and instrumental noise levels simultaneously.  The resulting ``detection significance grids" can then be used to guide the design of future experiments, especially as more data are collected and models refined.

\subsection{\label{ss:geometry}Cosmic Variance Limited Surveys}

Constraining the acoustic scale necessitates fine enough binning to resolve the BAO, implying a minimum survey size.  The BAO peaks are found near $k = \SI{0.1}{\h\per\mega\parsec}$ with width $\Delta k \sim \SI{0.06}{\h\per\mega\parsec}$. We therefore choose $k_{\text{min}} = \Delta k = \SI{0.02}{\h\per\mega\parsec}$, yielding a minimal survey dimension of $L = 2\pi / k_{\text{min}}  \sim $ \SI{300}{\mega\parsec\per\h}.  The peaks decay at $k \sim \SI{0.3}{\h\per\mega\parsec}$, so we consider scales $\SI{0.02}{\h\per\mega\parsec} < k < \SI{0.3}{\h\per\mega\parsec}$, consistent with galaxy surveys \cite{beutler17} and giving a maximum bin size of \SI{21}{\mega\parsec\per\h}.

\begin{figure}
\begin{center}
\includegraphics[width=1.0\linewidth]{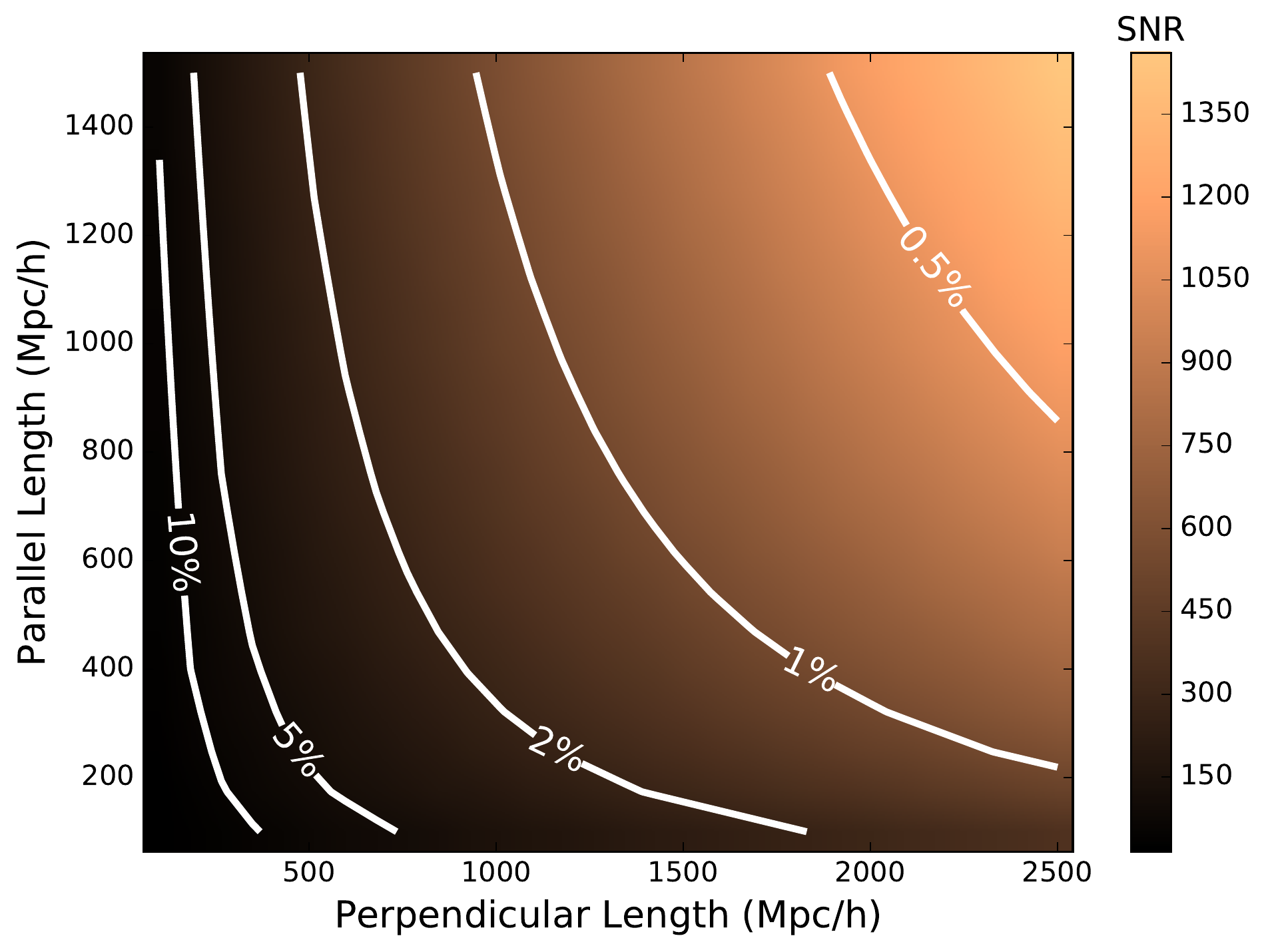}
\caption{Cosmic variance limits on the precision with which the BAO can be determined for \emph{noiseless} surveys as a function of parallel and perpendicular lengths, $L_\parallel$ and $L_\perp$. Contours show the achievable precision on the BAO scale, $\sigma(\alpha)$. 
The colormap shows the SNR on the total IM power spectrum (Eq.~\ref{eq:snr}).}
\label{fig:cv}
\end{center}
\end{figure}

A survey with the requisite volume to resolve the BAO does not necessarily sample enough modes to measure a compelling $\sigma(\alpha)$.  
To determine the cosmic variance limit \cite{knox95}, in Figure~\ref{fig:cv} we calculate the SNR for total power spectrum detection and $\sigma(\alpha)$ assuming a noiseless experiment while varying the survey volume.  Since the perpendicular ($L_{\perp}$) and transverse ($L_{\parallel}$) dimensions of survey volume are controlled by different aspects of the experiment---sky area and bandwidth, respectively---we vary both dimensions.  The total survey volume is $V_S = L_{\parallel}L_{\perp}^2$.  

The white contours show $\sigma(\alpha)$ levels from $0.5 - 10\%$ while the colormap indicates SNR. 
While 10\% constraints on $\sigma(\alpha)$ can be obtained with small survey volumes, $V_S \sim \SI{1.5e7}{(\mega\parsec\per\h)^3}$, the sample variance limit for competitive constraints pushes to significantly larger volumes---\SI{3.6e8}{(\mega\parsec\per\h)^3} for 2\% and  \SI{1.4e9}{(\mega\parsec\per\h)^3} for 1\%.  We also see that a detection could be made with $L_{\parallel}$ and $L_{\perp}$ that are highly unequal.  Finally, we note that high SNR on the power spectrum itself is needed for percent-level constraints on $\alpha$: a 5\% constraint requires $\text{SNR}\sim 100$ while 2\% requires $\text{SNR}\sim 275$.  

\subsection{\label{ss:sens}Noisy Surveys}

\begin{figure*}
\begin{center}
\includegraphics[width=0.8\linewidth]{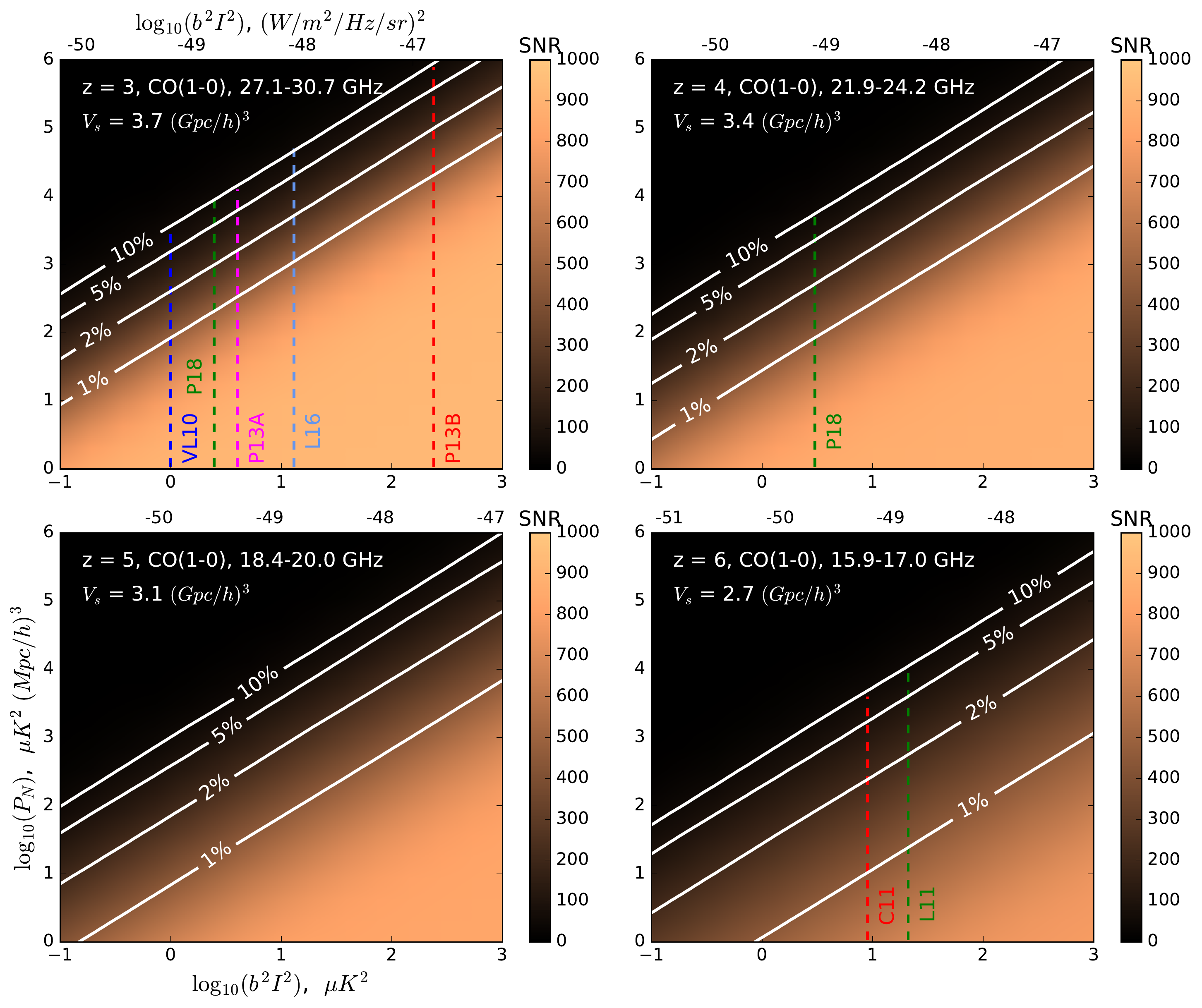}
\caption{CO power spectrum detection SNR (colormap) and  $\sigma(\alpha)$ (white contours) as a function of model amplitude $b^2 I^2$ ($x$-axis) and map noise level $P_N$ ($y$-axis).  Redshifts $z=3,4,5,6$ are shown in separate panels.  Various model predictions are overplotted as dashed, colored lines: V10 \cite{visbal10}, L16 \cite{li16}, P18 \cite{padmanabhan18}, P13B \cite{pullen13}, C11 \cite{carilli11}.  For models providing only a brightness temperature ($I$), we assumed a bias $b = z$ \cite{fonseca17}.  Model amplitudes in \si{\micro\kelvin^2} are converted to specific intensities in \si{(\watt\per\meter^2\per\hertz\per\steradian)^2} along the top axes; since the conversion is frequency-dependent, the scales change for each panel.}
\label{fig:COconts}
\end{center}
\end{figure*}

\begin{figure*}
\begin{center}
\includegraphics[width=0.8\linewidth]{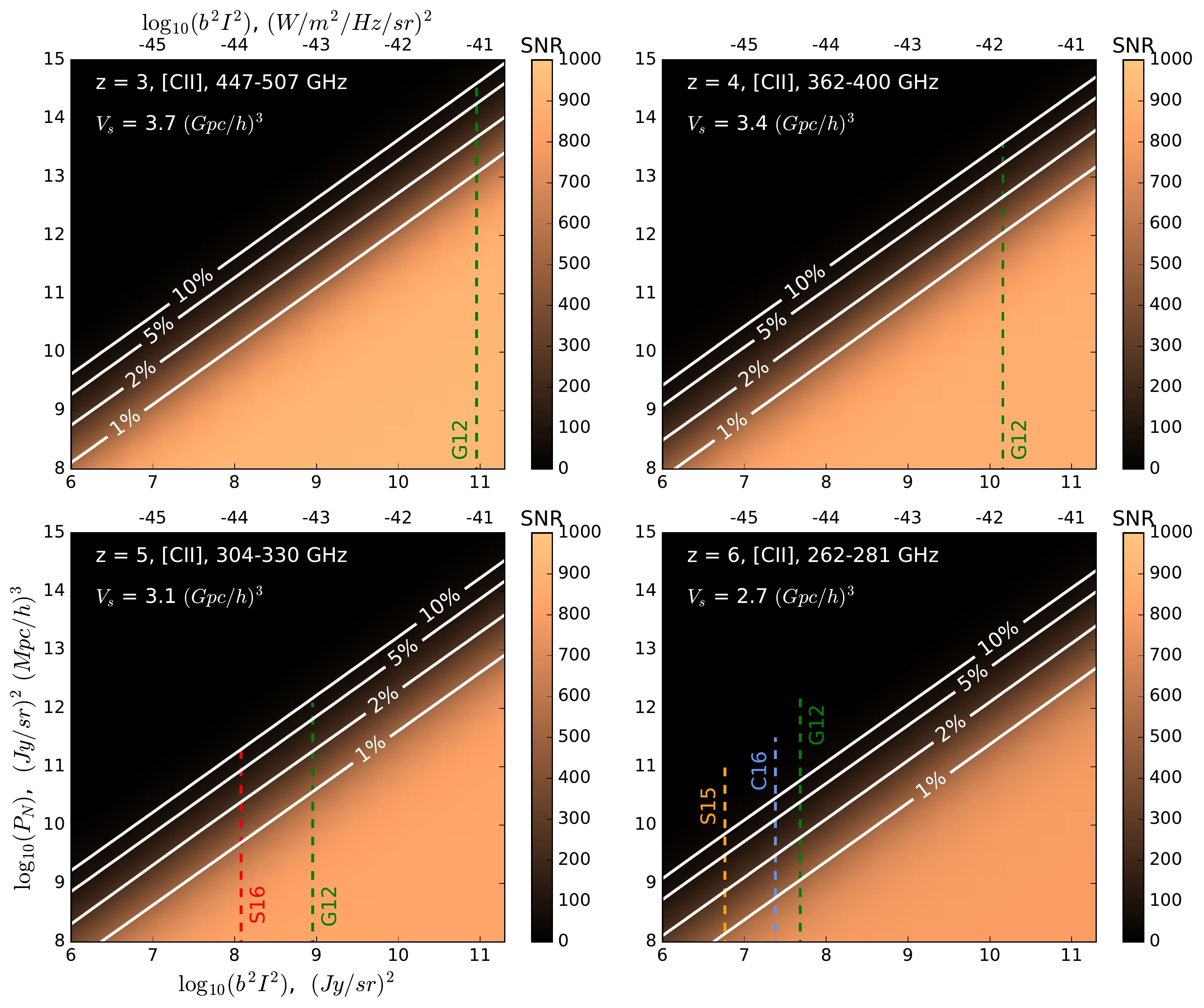}
\caption{[CII] power spectrum detection SNR (colormap) and $\sigma(\alpha)$ (white contours) as a function of model amplitude $b^2 I^2$ ($x$-axis) and map noise level $P_N$ ($y$-axis).  Redshifts $z=3,4,5,6$ are shown in separate panels.  Various model predictions are overplotted as dashed, colored lines: G12 \cite{gong12}, S15 \cite{silva15}, S16 \cite{serra16}, C16 \cite{cheng16}. For models providing only a brightness temperature ($I$), we assumed a bias $b = z$ \cite{fonseca17}.  Model amplitudes in \si{(\jy\per\steradian)^2} are converted to \si{(\watt\per\meter^2\per\hertz\per\steradian)^2} along the top axes.}
\label{fig:CIIconts}
\end{center}
\end{figure*}

Using the results derived in Section~\ref{ss:geometry}, we now calculate the map depths required to constrain the BAO in future  surveys; in Section~\ref{sec:future} we will evaluate how realistic it is to achieve such noise levels.  We consider measuring CO and [CII] with hypothetical $\SI{40}{\degree} \times \SI{40}{\degree}$ surveys, targeting redshifts $z = 3,4,5,6$ in bins of $\Delta z = 0.5$. Survey volumes thus range from \SI{3.7e9}{(\mega\parsec\per\h)^3} at $z=3$ to \SI{2.7e9}{(\mega\parsec\per\h)^3} at $z=6$ \footnote{Survey volume reduces with redshift because a fixed redshift bin width maps to smaller bandwidth at lower frequency; increasing bin size at higher redshift would counteract this at the expense of more evolution of the underlying signal.}. All of the volumes considered are large enough that sample variance does not dominate (up to $\sigma(\alpha) \sim \SI{1}{\percent}$).  We assume a \SI{12}{\meter} aperture and \SI{10}{\mega\hertz} spectral resolution for CO, and a \SI{10}{\meter} aperture with \SI{400}{\mega\hertz} spectral resolution for [CII], as planned in near-future experiments \cite{li16, serra16, silva15}.

Figure~\ref{fig:COconts} shows the SNR on the total CO(1-0) power spectrum (colormap) and $\sigma(\alpha)$ (contours) as both the power spectrum amplitude and noise level are varied.  
The model amplitude is parametrized as $b^2 I^2$ in units of \si{\micro\kelvin^2}, while the noise $P_N$ is parametrized as the variance (\si{\micro\kelvin^2}) in \SI{1}{(\mega\parsec\per\h)^3} pixels. Figure~\ref{fig:CIIconts} shows the same detection statistics for [CII].  In this case, for comparison to the literature, $b^2 I^2$ is in units of \si{(\jy\per\steradian)^2} and $P_N$ is in \si{(\jy\per\steradian)^2(\mega\parsec\per\h)^3}.  Both scales are additionally converted to common \si{(\watt\per\meter^2\per\hertz\per\steradian)^2} specific intensities along the top axes.  To illustrate the spread in model predictions, in both plots we overlay a representative sample of $b^2 I^2$ models from the literature as vertical lines \footnote{For several literature models only a brightness temperature $I$ was provided; in these cases we assumed a bias of $b = z$ as motivated by Ref.~\cite{fonseca17}.}.  For a given model, the map noise level required to measure $\sigma(\alpha)$ to some level is given by the $P_N$ value of the intersection between the model line and that contour.

In general it is more difficult to detect the BAO at higher redshift, even if $b^2 I^2$ is constant across redshift.  This is due both to $P_{\delta\delta}$ decreasing at higher $z$ and beam smoothing, which becomes more significant at lower frequencies for a fixed aperture size.  It is also worth noting that noise levels are not directly comparable across redshift bins because of the different physical extents of pixel sizes:  a \SI{1}{(\mega\parsec\per\h)^2} patch subtends \SI{0.51}{\arcmin^2} at $z=3$ but only \SI{0.30}{\arcmin^2} at $z=6$.

In these results we have not included shot noise in the power spectrum model.  At the lower end of redshifts considered here it is expected to be subdominant to the clustering signal at the scales of interest for the BAO.  Ref.~\cite{li16}, for example, predict a $z=3$ shot noise amplitude that is $< \SI{5}{\percent}$ of the clustering amplitude at $k = \SI{0.1}{\h\per\mega\parsec}$.  At earlier times this may be less accurate: Ref.~\cite{dizgah18} suggests that at $z = 6$, $P_{\text{shot}} = P_{\delta\delta}$ at $k \sim \SI{0.2}{\h\per\mega\parsec}$. In a noiseless experiment, this would inflate the uncertainties by a factor of $\sqrt{2}$.  However, for current experiments targeting the clustering signal it is unlikely that shot noise will dominate the error budget, and compared to the large spread in model predictions shown in Figures~\ref{fig:COconts} and \ref{fig:CIIconts} it represents a small part of our uncertainty in the noise levels required to constrain the BAO. 

\section{\label{sec:future}Prospects for Future Experiments}

In this section, we evaluate the feasibility of realistic future experiments to reach the sensitivity necessary for useful constraints on the BAO scale.
As the detection SNR for a particular experiment is dependent on the expected signal (which may vary by orders of magnitude), we express experimental sensitivity in terms of \emph{survey weight}, $\text{SW} = V_S / P_N$. Here $V_S$ is the survey volume and $P_N$ is the map noise, or rms fluctuation in a standardized IM voxel.
Survey weight, expressed in $\si{\micro\kelvin^{-2}}$ or $\si{(\jy\per\steradian)^{-2}}$, scales linearly with integration time, detector count, and sensitivity to the power spectrum, enabling easy comparison of one experiment to another.  We consider CO(1-0) at $z=3$ and [CII] at $z=6$ since they correspond to experiments that are already underway and there are several model predictions in the literature at these redshifts. 

The COMAP experiment, currently in the commissioning phase, is targeting CO(1-0) at $z=3$.  Since the initial experiment is not expected to detect the power spectrum at the high SNRs ($\gtrsim 100$) required to constrain $\alpha$, we consider a second-generation ``COMAP full" experiment (described in Ref~\cite{li16}) consisting of 500 dual-polarization feeds with $T_{\text{sys}} = \SI{35}{\kelvin}$ and \SI{10}{\mega\hertz} channels.  We consider a measurement from \num{27.1}--\SI{30.7}{\giga\hertz} corresponding to $\Delta z = 0.5$.  Measuring a model in the middle of the distribution of the top left panel of Figure~\ref{fig:COconts} (here we choose the ``L16" model \cite{li16}, with $b^2I^2 \approx \SI{13}{\micro\kelvin^2}$) to $\sigma(\alpha) = \SI{5}{\percent}$ requires $P_N = \SI{2e4}{\micro\kelvin^2.(\mega\parsec\per\h)^3}$ over a survey volume of $V_S = \SI{3.7e9}{(\mega\parsec\per\h)^3}$, or a total survey weight of \SI{1.9e5}{\micro\kelvin^{-2}}.  This would require $\SI{15700}{\hour}$ integration time.  It is worth noting that a recent tentative detection of CO power at this redshift \cite{keating16} is a factor of $\sim 4$ higher than the L16 model (however, it is only sensitive to shot noise at high $k$).  If that factor also applies to the clustering regime, the $5\%$ measurement could be made in \SI{3900}{\hour} and a $2\%$ constraint would take \SI{12200}{\hour}.  Furthermore, COMAP is using \SI{8}{\giga\hertz} of bandwidth, enabling measurement of several redshift bins simultaneously.  

At the other end of the redshift range, we look at [CII] at $z = 6$ which is targeted by TIME \cite{crites14} and CONCERTO \cite{serra16}.  Again, they are not expected to achieve the requisite SNR for BAO constraints.  We therefore consider the ``CII-Stage II" survey in \cite{silva15}, which consists of a 64-spectrometer instrument with total Noise Equivalent Flux Density of \SI{5}{\milli\jy.\sqrt{\second}} and \SI{400}{\mega\hertz} spectral resolution.  For the $z=6$ measurement we consider \num{262}--\SI{281}{\giga\hertz}.  Choosing a model in the middle of the distribution, we find that measuring the ``C16" model \cite{cheng16} ($b^2I^2 \approx \SI{2.4e7}{(\jy\per\steradian)^2}$) to $\sigma(\alpha) = \SI{5}{\percent}$ requires $P_N = \SI{1e10}{(\jy\per\steradian)^2.(\mega\parsec\per\h)^3}$ over $V_S = \SI{2.7e9}{(\mega\parsec\per\h)^3}$, or a total survey weight of \SI{2.7e-1}{(\jy\per\steradian)^{-2}}.  This would require $\SI{2250}{\hour}$ of integration time; a 2\% measurement would take $\SI{7500}{\hour}$.  Again, the [CII] experiments also use wide bandwidths so several redshift bins are measurable.  The tentative detection of Ref.~\cite{pullen18} is also promising, as it favors more optimistic models (albeit at $z \sim 2.6$).  

In both cases, the calculated integration times (in the several- to ten-thousand hour range) are reasonable for dedicated, next-generation multiyear surveys.

\section{\label{sec:cosmology}Implications for Cosmology}

We have shown that plausible near-future IM experiments can---provided the underlying line power spectra are near or brighter than the median of current predictions---make precision measurements of the BAO scale in reasonable integration times and constrain the $3 < z < 6$ expansion history at a level of $5\%$ or better. 
We can then estimate constraints on $w_0$ and $w_a$ using 
\begin{equation}
\sigma(w) = \frac{\sigma(H)}{dH/dw}
\end{equation}
and $\sigma(H) = H(z) \,\sigma(\alpha)$.
Assuming a standard dark energy equation of state, $w = w_0 + (1-a) w_a$, $\Omega_\Lambda = 0.7$, $\Omega_m = 0.3$ and $H_0 = \SI{70}{\kilo\meter\per\second\per\mega\parsec}$, a $5\% \ (2\%)$ measurement would constrain $w_0 \approx -1 \pm 0.7 \ (0.3)$ and $w_a \approx \pm 0.9 \ (0.4)$.  The best constraints are from the $z=3$ bin, while limits at $z=6$ are a factor of $\sim 3.5$ worse. These relatively uncompetitive constraints are expected; the constraining power of the BAO drops sharply at $z > 2.4$ because the Universe is highly matter dominated. Nevertheless, such limits are still valuable because IM experiments can probe hitherto unexplored redshifts.

To better evaluate the constraining power of IM experiments on more exotic dark energy models, we consider the early dark energy model of Ref.~\cite{Karwal2016}. In this model, an early dark energy component with density $\Omega_{\text{ede}}$ approximates a cosmological constant with $w \approx -1$ at high redshift and as a scalar field with $w = 1$ at low redshift. We define a critical redshift $z_c$ at which the transition between cosmological constant and scalar field occurs. All allowed models have $z_c \gg 10$, so for evaluating constraints from IM experiments from $z = 3-6$ we can approximate the energy density of this component as $\Omega_{\text{ede}}(a) = \Omega_{\text{ede}}^0 a^{-6}$. A $\sigma(\alpha) = 0.05 \ (0.02)$ constraint leads to $\sigma(\Omega_{\text{ede}}^0) \approx 9 \ (4) \times 10^{-4}  $ at $z=6$ and $\sigma(\Omega_{\text{ede}}^0) \approx 5 \ (2) \times 10^{-4}$ at $z=3$, similar to the current CMB limits quoted by Ref.~\cite{Karwal2016}.

Finally, the most general model for dark energy is to reconstruct the expansion history using a spline~\cite{Bernal2016}. 
Ref.~\cite{Poulin2018} performed such a reconstruction and found that the Lyman-$\alpha$ forest BAO measurement led to a preference for an expansion rate $5\%$ lower than $\Lambda$CDM at $z=2.5$, which a $5\%$ measurement of $\alpha$ at $z=3-6$ would confirm or rule out.

\section{\label{sec:conclusion}Conclusions}

We have investigated the potential for a high-redshift BAO measurement from line intensity mapping experiments, in particular CO and [CII] emission lines. We find that realistic near-future experiments such as COMAP ``full" or CII-Stage II may be able to constrain the expansion rate at the $5\%$ level, contingent on the signal being close to the median theoretical models. 
Because IM probes uniquely high redshift ranges, even a first detection by these experiments would already be able to place competitive constraints on exotic and early dark energy models.
Furthermore, these surveys are strongly limited by experimental noise; successor surveys of similar volume closer to the cosmic variance limit could potentially reach a $<1\%$ constraint on the expansion rate. 

Our simple estimates are highly dependent on which theoretical model for the IM signal is assumed. The highest brightness temperatures allowed by theory would provide better constraints on the expansion scale, while the lowest theoretical models would make a BAO detection challenging.

Many considerations that are not included in our simple estimates will factor into real experiments, the most prominent being foregrounds.  
CO experiments will have to contend with Galactic synchrotron emission \cite{planck_synch}, especially over the large sky areas needed, and potentially anomalous microwave emission \cite{leitch97}. HI experiments are now developing techniques for removing the smooth-spectrum synchrotron signal at lower frequencies where the signal is much brighter \cite{liu09}; the same methods should be applicable to CO.  For [CII], interloper lines from other atomic and molecular species at different redshifts will need to be removed.  Work is now ongoing to develop line foreground removal methods \cite{lidz16, cheng16, sun18}.  At the same time, these interloper lines represent an opportunity for additional constraints; for example, the higher-$J$ transition CO(8-7) from $3 < z < 4$ is measurable by the high-$z$ [CII] experiment considered above.  

In this paper we have only considered an isotropic, spherically-averaged power spectrum.  But since both smooth-spectrum and line foregrounds affect the transverse and line-of-sight components of the power spectrum differently, in future work it will be worthwhile to analyze errors on the two-dimensional power spectrum and break up our one-dimensional $\alpha$ constraints into $\alpha_{\parallel}$ and $\alpha_{\perp}$, which independently probe the Hubble rate and angular diameter distance, respectively \cite{bull15}.

We have focused on CO and [CII] due to their use in existing and near-future instruments.  However, many other candidate lines exist: HI at frequencies between EoR experiments and lower-redshift BAO experiments, Ly$\alpha$/H$\alpha$ (e.g. from space with SPHEREX \cite{dore14} and in conjunction with the CIB at higher redshift \cite{kashlinsky15}), and many far infrared lines such as [NII] and [OI] \cite{uzgil14, serra16}. These lines are also excellent candidates for probing the BAO and we expect that future work will explore the noise levels required for their detection.

\begin{acknowledgments}
We would like to thank Pete Barry, Abby Crites, Wayne Hu, and Erik Shirokoff for helpful conversations.  KSK acknowledges support from the Grainger Foundation and the Kavli Institute for Cosmological Physics at the University of Chicago through an endowment from the Kavli Foundation and its founder Fred Kavli.
\end{acknowledgments}

\bibliography{apsamp}
\end{document}